\documentclass[useAMS,usenatbib]{mnras}
\usepackage{natbib}
\usepackage{epsfig}
\usepackage{scrtime}
\usepackage{graphicx}   
\usepackage{amsmath}    
\usepackage{amssymb}    
\usepackage{float}
\usepackage[caption = false]{subfig}
\usepackage{multirow}
\usepackage{rotating}
\usepackage{ulem}
\usepackage{booktabs}

\usepackage{xcolor}
\usepackage{soul}

\newcommand{\mincir}{\raise-3.truept\hbox{\rlap{\hbox{$\sim$}}\raise4.truept\hbox{$<$}\ }}

\usepackage[T1]{fontenc}
\usepackage{ae,aecompl}

\usepackage{newtxtext,newtxmath}

\title[Simulations of HuBi1]{Formation and fate of the born-again planetary nebula HuBi\,1}

\author[Toal\'{a} et al.]{J.A.\,Toal\'{a}$^{1}$\thanks{E-mail: j.toala@irya.unam.mx}, V.\,Lora$^{1}$, B.\,Montoro-Molina$^{2}$,  M.A.\,Guerrero$^{2}$ and A.\,Esquivel$^{3}$ \\
$^{1}$Instituto de Radioastronom\'ia y Astrof\'isica, UNAM, Campus Morelia, C.P. 58089, Morelia, M\'exico\\
$^{2}$Instituto de Astrof\'{i}sica de Andaluc\'{i}a, IAA-CSIC, Glorieta de la Astronom\'{i}a S/N, Granada 18008, Spain\\
$^{3}$Instituto de Ciencias Nucleares, UNAM, A. P. 73-543, 04510, Ciudad de M\'{e}xico, Mexico
}

\begin{document}
\maketitle
\label{firstpage}
\begin{abstract}


\noindent 
We present the first 3D radiation-hydrodynamic simulations on the formation and evolution of born-again planetary nebulae (PNe), with particular emphasis to the case of HuBi\,1, the inside-out PN. 
We use the extensively-tested {\sc guacho} code to simulate the formation of HuBi\,1 adopting mass-loss and stellar wind terminal velocity estimates obtained from observations presented by our group. 
We found that, if the inner shell of HuBi\,1 was formed by an explosive {\it very late thermal pulse} (VLTP) ejecting material with velocities of $\sim$300 km~s$^{-1}$, the age of this structure is consistent with that of $\simeq$200 yr derived from multi-epoch narrow-band imaging. Our simulations predict that, as a consequence of the dramatic reduction of the stellar wind velocity and photon ionizing flux during the VLTP, the velocity and pressure structure of the outer H-rich nebula are affected creating turbulent ionized structures surrounding the inner shell. These are indeed detected in Gran Telescopio Canarias MEGARA optical observations. 
Furthermore, we demonstrate that the current relatively low ionizing photon flux from the central star of HuBi\,1 is not able to completely ionize the inner shell, which favors previous suggestions that its excitation is dominated by shocks. 
Our simulations suggest that the kinetic energy of the H-poor ejecta of HuBi\,1 is at least 30 times that of the clumps and filaments in the evolved born-again PNe A\,30 and A\,78, making it a truly unique VLTP event.

\end{abstract}

\begin{keywords}
  stars: evolution --- stars: low-mass --- stars: mass-loss --- stars: AGB and post-AGB ---
  (ISM:) planetary nebulae: general --- (ISM:) planetary nebulae: individual (HuBi\,1) 
\end{keywords}

\section{Introduction}

HuBi\,1 is part of the selected group of planetary nebulae (PNe) named
born-again PNe that are thought to have experienced a {\it very late
  thermal pulse} \citep[VLTP; e.g.,][]{Schonberner1979,Iben1983}.
During this specific evolutionary phase, the He-burning shell at the
surface of the central star of a PN (CSPN) reaches the conditions to
ignite He into C and O through an explosive event.  This thermonuclear
event ejects H-deficient and C-rich material inside the old H-rich PN
\citep[see, e.g.,][and references therein]{MB2006} rendering the star
as a C-rich [Wolf-Rayet] ([WR]) type star \citep{GT2000}.

Recent works have presented stark evidence of the dramatic changes
experienced by the CSPN of born-again PNe \citep[see, e.g., the case
  of the Sakurai's Object;][]{Evans2020}, and in particular HuBi\,1.
\citet{Guerrero2018} demonstrated that in less than 50~yr its CSPN
declined its brightness in about 10~mag changing its atmosphere and,
as a consequence, producing changes in the ionization structure of its
surrounding PN.

\begin{figure*}
\centering
\includegraphics[width=0.9\linewidth]{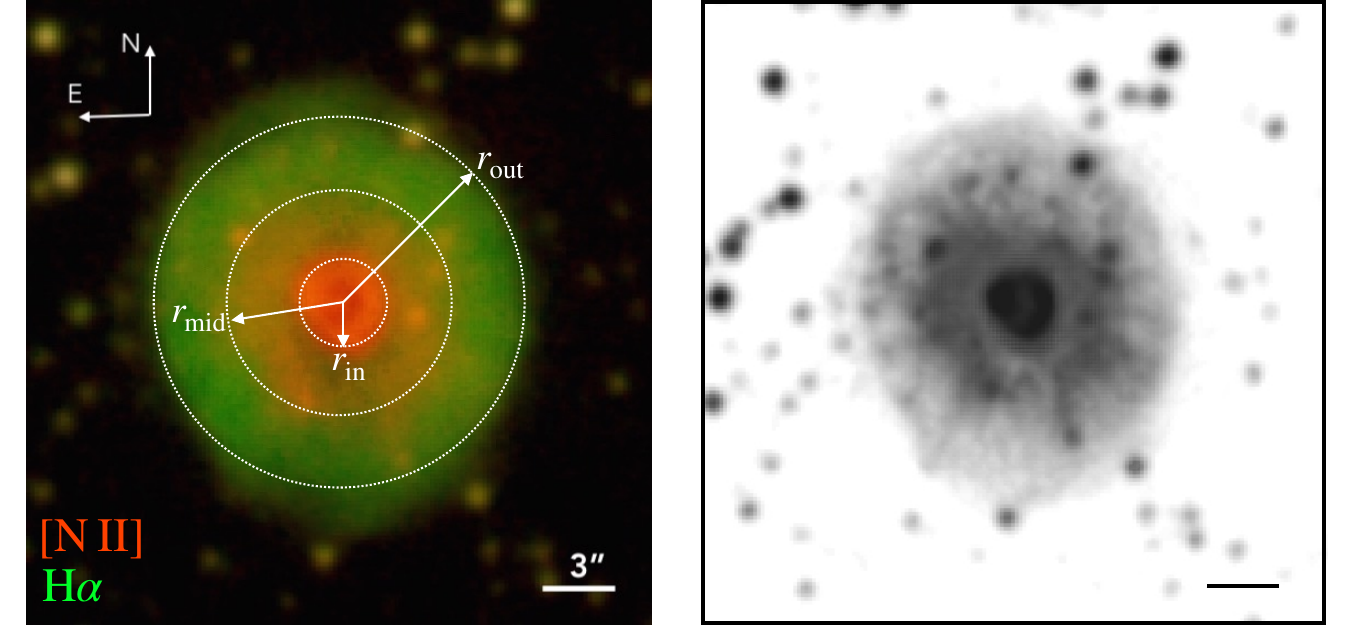}
\caption{Optical images of HuBi\,1. Left: Colour-composite optical
  image of HuBi obtained with the [N\,{\sc ii}] (red) and H$\alpha$
  (green) narrow-band filters at the NOT. The outer
  ($r_\mathrm{out}$), middle ($r_\mathrm{mid}$) and inner
  ($r_\mathrm{in}$) radii have extension of 8, 5 and 2~arcsec,
  respectively. The image was adapted from \citet{Rechy2020}. Right:
  Grey-scaled [N\,{\sc ii}] narrow-band image. Both panels have the
  same FoV. North is up, east to the left.}
\label{fig:opt}
\end{figure*}

HuBi\,1 has a double-shell structure (see Fig.~\ref{fig:opt}). The
outer shell with an angular radius of $r\sim8$~arcsec is dominated by
emission from recombination lines of H~{\sc i} and He~{\sc i}, whilst
its inner shell with $r\sim2$~arcsec is dominated by emission from
forbidden lines \citep{Guerrero2018}.  The inner shell has a notable
inverted ionization structure, with the emission from higher
ionization species such as O$^{++}$ and He$^{++}$ encompassing that
from lower ionization species such as N$^{+}$, O$^{+}$ and S$^{+}$
\citep[][Montoro-Molina et al., in preparation]{Guerrero2018}.  Such
unusual inverted ionization structure gave HuBi\,1 the title of {\it
  inside-out} PN.

\citet{Guerrero2018} used multi-epoch observations of HuBi\,1,
state-of-the-art stellar atmosphere models of its CSPN from the {\sc
  powr} code \citep[see][and references
  therein]{Sander2015}\footnote{\url{http://www.astro.physik.uni-potsdam.de/~wrh/PoWR/powrgrid1.php}}
and modern stellar evolution models from \citet{MB2016} to predict
different aspects of the evolution of this PN and its progenitor
star. In particular, these authors found that a model that experienced
a mass-loss rate during the VLTP ($\dot{M}_\mathrm{VLTP}$) of
7.6$\times10^{-5}$~M$_\odot$~yr$^{-1}$ fits the evolutionary status of
the CSPN of HuBi\,1, with an ejected mass during the VLTP
($M_\mathrm{VLTP}$) of 8$\times10^{-4}$~M$_\odot$.
\citet{Guerrero2018} estimated that currently the wind velocity of the
CSPN has a velocity of 360~km~s$^{-1}$ with a mass-loss rate of
$8\times10^{-7}$~M$_\odot$~yr$^{-1}$.  These authors also estimated a
relatively low ionizing photon flux of $\approx10^{44}$~s$^{-1}$,
which led them to propose that the outer shell is recombining.

In \citet{Rechy2020} we have studied HuBi\,1 using integral-field
spectroscopic Multi-Espectr\'ografo en GTC de Alta Resoluci\'on para
Astronom\'\i a \citep[MEGARA;][]{Gil2018} mounted on the Gran
Telescopio Canarias (GTC).  The unrivaled tomographic capability of
these MEGARA observations have unveiled the kinematic signature of the
inner shell in HuBi\,1, revealing that it was ejected about 200~yr ago
and currently has an expansion velocity of $\approx$300 km~s$^{-1}$.

The MEGARA observations showed that the inner structure is apparently
distributed in a shell-like morphology, very different to what is
observed in other born-again PNe.  For example, {\it Hubble Space
  Telescope} ({\it HST}) and IR observations of A\,30, A\,58 and A\,78
have revealed that the H-deficient material ejected during the VLTP
has a marked bipolar morphology.  The material in the born-again
ejecta of these PNe is distributed in a disk-like structure and a
bipolar ejection that resembles a jet
\citep[][]{Borkowski1993,Borkowski1995,Clayton2013,Fang2014}.  In
particular, for the cases of the more evolved born-again PN A\,30 and
A\,78, the ring-like structure appears to have been disrupted by the
complex interactions with the stellar wind and ionizing photon flux
from their CSPNe \citep[see][and references therein]{Toala2021}.

Our group presented the first attempt to model the formation of
born-again PNe in \citet{Fang2014}.  In that work we presented 2D
radiation-hydrodynamic numerical simulations tailored to the
born-again PNe A\,30 and A\,78 in comparison with a study of the
expansion of their H-deficient clumps and filaments inside their old
PNe. Those simulations demonstrated that adopting a velocity of
20~km~s$^{-1}$ during the VLTP can help explaining the distribution of
the H-poor knots, the dynamical age ($\sim$1000~yr) and evolution of
A\,30 and A\,78. Our simulations showed that the C-rich material will
be disrupted by a combination of effects. Hydrodynamical
instabilities, mainly Rayleigh-Taylor will break the VLTP material
into clumps and filaments.  These will be subsequently ionized and
photoevaporated by the increasing UV flux from the CSPN.  Finally, the
current fast stellar wind will also play a role in dragging the
material with the denser and slower clumps remaining close to the
CSPN.

In this paper we present 3D radiation-hydrodynamic numerical
simulations of the formation and evolution of born-again PNe, with
emphasis to HuBi\,1.  The simulations are use to explain the formation
of HuBi\,1 and to peer into its further evolution.  This is assessed
by adopting different initial conditions for the VLPT ejecta in the
simulations (2 cases are explored).  A comparison with more evolved
born-again PNe is also attempted.

This paper is organized as follows. In Section~2 we present the code
used to run our simulations and describe the initial
conditions. Section~3 describes the different numerical results
obtained from the simulations. A discussion is presented in Section~4
and a summary of the work is presented in Section~5.

\section{Simulations}

We used the extensively-tested radiation-hydrodynamic 3D code {\sc
  guacho} \citep{Esquivel2009,Esquivel2013} to model the formation and
evolution of the born-again PN HuBi\,1.  {\sc guacho} includes a
modified version of the ionizing radiation transfer presented in
\citet{Raga2009}.  It solves the gas-dynamic equations with a second
order accurate Godunov-type method, using a linear slope-limited
reconstruction and the HLLC approximate Riemann solver
\citep{Toro1994} implemented on a uniform Cartesian grid.

Simultaneously with the Euler equations, we solve the rate equation
for neutral and ionized hydrogen
\begin{equation}
    \frac{\partial n_\mathrm{HI}}{\partial t} + \nabla \cdot(n_\mathrm{HI} {\bf u})= 
    n_\mathrm{e} n_\mathrm{HII} \alpha(T) - n_\mathrm{HI} n_\mathrm{HII} c(T) - n_\mathrm{HI} \phi,
\end{equation}
\noindent where {\bf u} is the flow velocity, $n_\mathrm{e}$,
$n_\mathrm{HI}$ and $n_\mathrm{HII}$ are the electron, neutral
hydrogen and ionized hydrogen number densities, $\alpha(T)$ is the
recombination coefficient, $c(T)$ is the the collisional ionization
coefficient and $\phi$ is the H photoionization rate due to a central
source. The photoionizing rate is computed with a Monte-Carlo ray
tracing method, described in \citet{Esquivel2013} and
\citet{Schneiter2016}.

We define the ionization fraction as
\begin{equation}
    \chi=\frac{n_\mathrm{HII}}{n_\mathrm{HI}+n_\mathrm{HII}},
\end{equation}
\noindent with the total number density defined as
$n=n_\mathrm{HI}+n_\mathrm{HII}$.  The ionization fraction is used to
estimate the radiative cooling, which is added to the energy equation
using the prescription described in \citet{Esquivel2013}.

The simulations presented here have been performed on a 3D cartesian
grid with a resolution of $(x, y, z)$=(600, 600, 600) on a box of
(0.6$\times$0.6$\times$0.6)~pc$^{3}$ in physical size, that is, a cell
resolution of 0.001~pc. The injection cells correspond to the
innermost 0.01~pc in the simulation.

\begin{figure}
\centering
\includegraphics[width=0.9\linewidth]{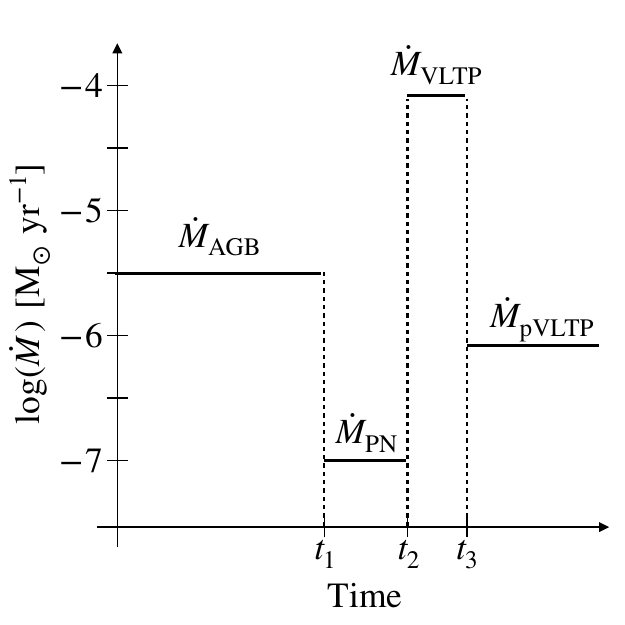}\\
\includegraphics[width=0.9\linewidth]{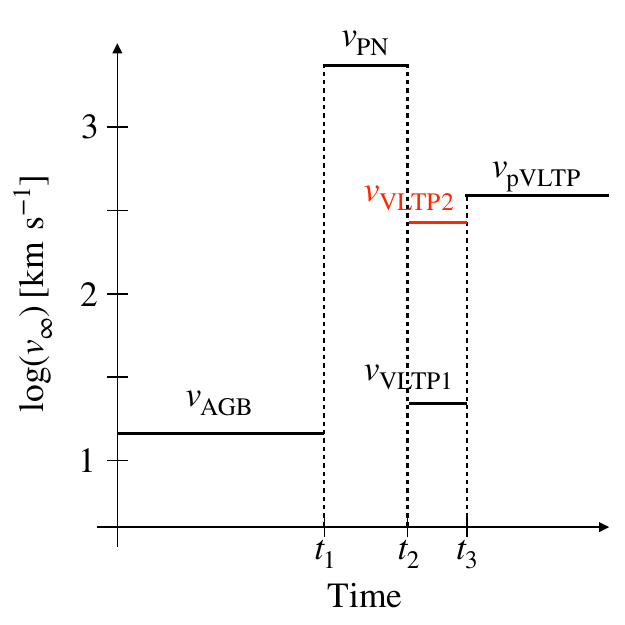}
\caption{Illustration of the evolution with time of the mass-loss rate
  ({\bf top} panel) and stellar wind velocity ({\bf bottom} panel) of
  the simulations used here. The different phases are labeled. Note
  that there are two values for the velocity on during the VLTP
  ($v_\mathrm{VLTP}$) phase which correspond to the two simulations
  presented here.}
\label{fig:hist}
\end{figure}

\subsection{Initial conditions - old PN formation}

We started with an homogeneous ISM with initial density and
temperatures of $n_{0}$=1~cm$^{-3}$ and $T_{0}=$100~K,
respectively. We first model the creation of the old H-rich PN of
HuBi\,1.  For this, we first launch a slow wind corresponding to the
AGB stage with mass-loss rate
$\dot{M}_\mathrm{AGB}=10^{-5.5}$~M$_\odot$~yr$^{-1}$ and a velocity of
$v_\mathrm{AGB}=$15~km~s$^{-1}$ for a total time of
$t_{1}=5\times10^5$~yr.  This creates a density distribution with a
dependence with radius of $n \sim r^{-2}$ as commonly obtained for the
AGB phase \citep[see, e.g.,][]{Villaver2002}. No photoionization flux
is included during this phase.

\begin{table}
  \begin{center}
     \caption{Imput parameters for the different phases in our simulations.}
     \begin{tabular}{llcccc}
     \hline
  & Phase & $\dot{M}$                & $v_{\infty}$   & duration  & Note\\
  &       & (M$_{\odot}$~yr$^{-1}$)  & (km~s$^{-1})$  & (yr)      &     \\
     \hline
$\;\;\;\, \rightarrow t_1$ & AGB      & $10^{-5}$             & 15             & $5\times10^{5}$ & Run A and B\\
$t_1 \rightarrow t_2$ & post-AGB & $10^{-7}$             & 2000           & 2300            & Run A and B\\
$t_2 \rightarrow t_3$ & VLTP1    & $7.6\times10^{-5}$    & 20             & 20              & Run A\\
$t_2 \rightarrow t_3$ & VLTP2    & $7.6\times10^{-5}$    & 300            & 20              & Run B\\
$t_3 \rightarrow$     & pVLTP    & $8.0\times10^{-7}$    & 360            &                 & Run A and B\\ 
\hline 
     \end{tabular}
     \label{tab:parameters}
  \end{center}
\end{table}

\begin{figure}
\centering
\includegraphics[width=\linewidth]{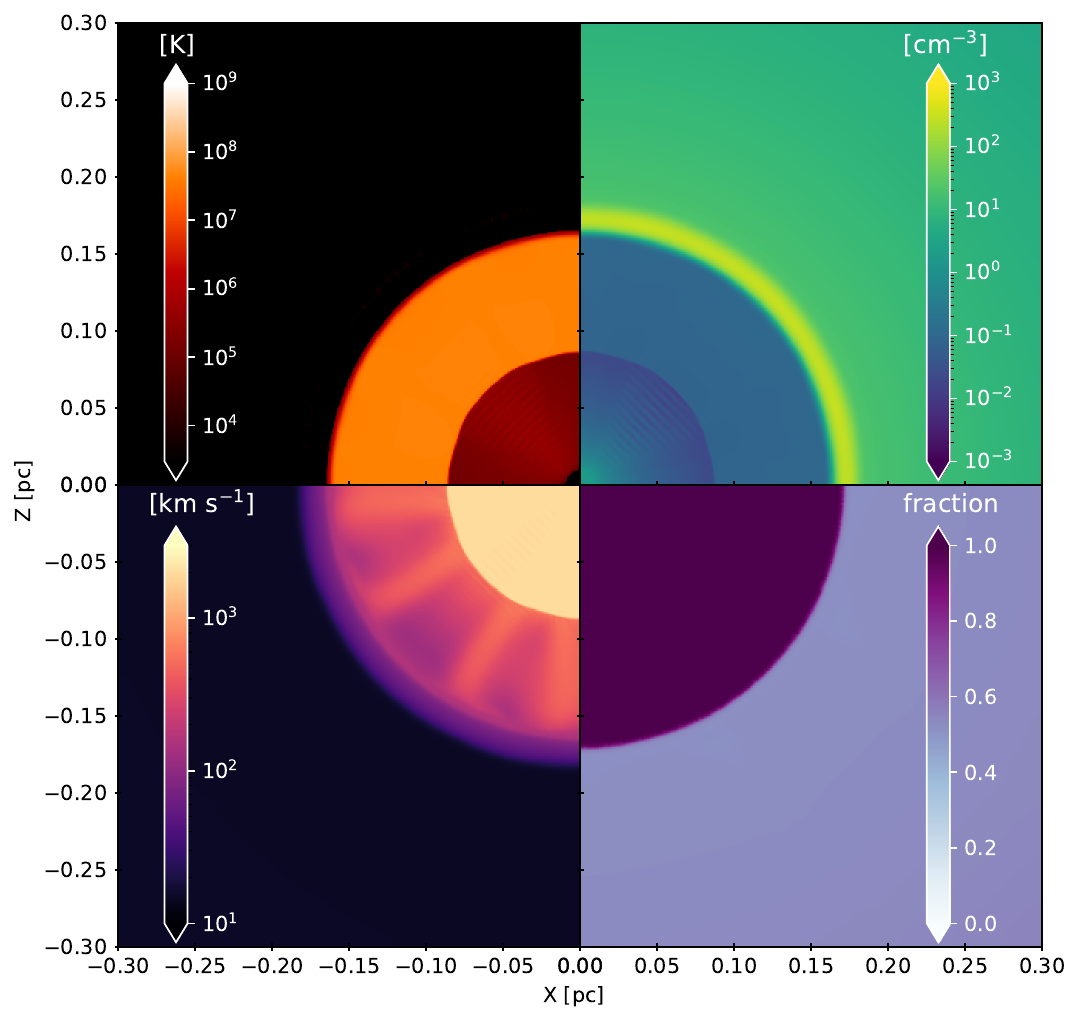}
\caption{ Total number density ($n$ - top right), temperature ($T$ -
  top left), expansion velocity ($v$ - bottom left) and ionization
  fraction ($\chi$ - bottom right) for the simulation at $t=t_2$.
  This represents the initial conditions (the old PN) just before the
  onset of the VLTP phase.  }
\label{fig:initial_c}
\end{figure}

Secondly, a post-AGB phase which creates the old PN is modeled by
injecting a fast wind with a velocity of
$v_\mathrm{PN}$=2000~km~s$^{-1}$ and a mass-loss rate of
$\dot{M}_\mathrm{PN}=10^{-7}$~M$_\odot$~yr$^{-1}$.  An ionizing photon
flux of $5\times10^{46}$~s$^{-1}$ is adopted for this
phase. Figure~\ref{fig:hist} shows an illustration of the evolution
with time of the stellar wind parameters in these two phases.

The simulation is run until the shell of the PN reaches a radius of
$\sim$0.18~pc so that by the time the born-again event occurs it could
reach a 0.2~pc similarly to what is currently observed in
HuBi\,1\footnote{The angular radius of HuBi\,1 is $\sim$8~arcsec which
  is $\sim$0.2~pc (see Fig.~\ref{fig:opt}) at a distance of 5.3~kpc
  \citep{Frew2016}.}. The formation of the old H-rich PN occurs during
$t_1$ and $t_2$ in Figure~\ref{fig:hist}.

\begin{figure*}
\centering
\includegraphics[width=0.47\linewidth]{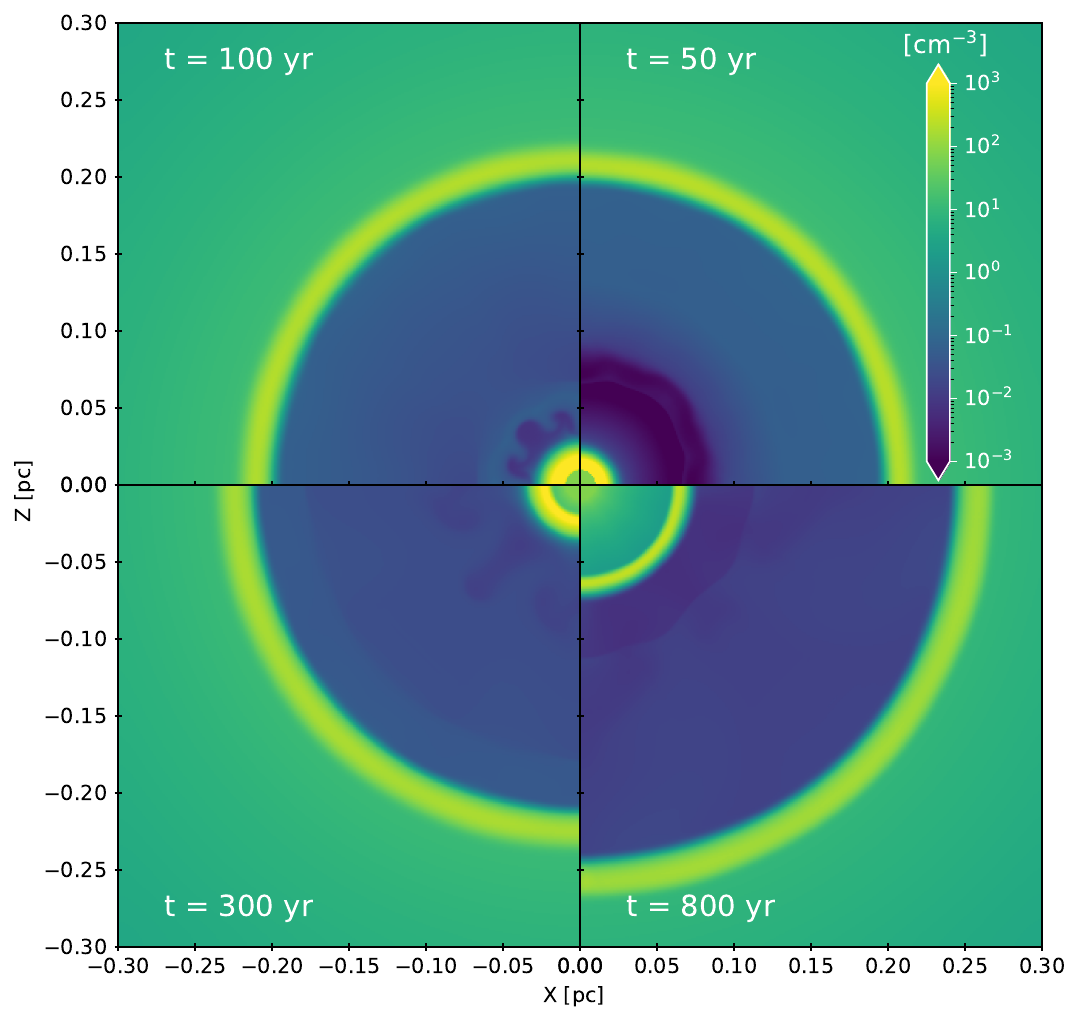}
\includegraphics[width=0.47\linewidth]{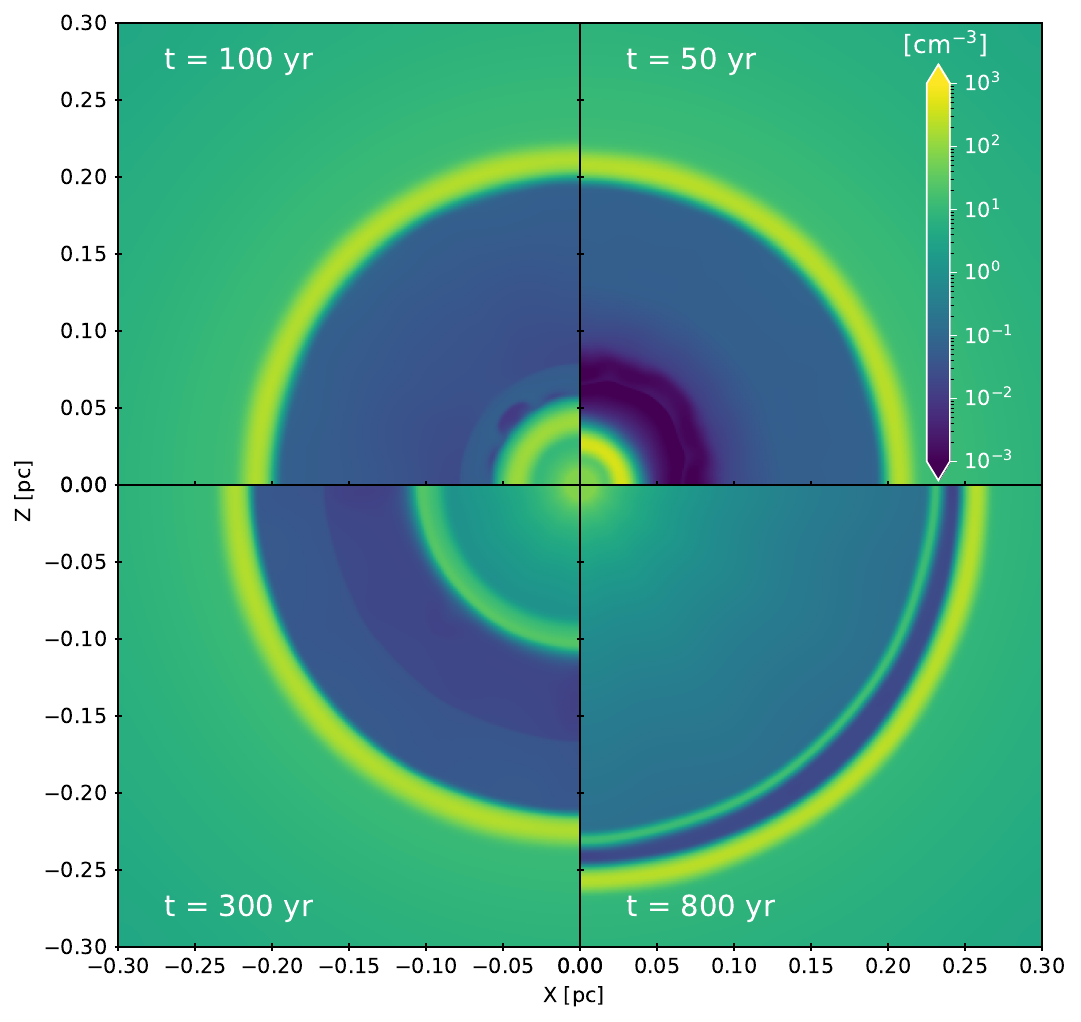}
\caption{Number density $n$ in the $x-z$ plane ($y$=0) of the two
  simulations presented here.  The left panel corresponds to Run~A
  ($v_\mathrm{VLTP1}$=20~km~s$^{-1}$) and the right panel to Run~B
  ($v_\mathrm{VLTP2}$=300~km~s$^{-1}$). The sub-panels show different
  time steps with $t=0$ marking the onset of the pVLTP phase, that is,
  at $t=t_3$ in Figure~\ref{fig:hist}.}
\label{fig:rho}
\end{figure*}

\begin{figure*}
\centering
\includegraphics[width=0.47\linewidth]{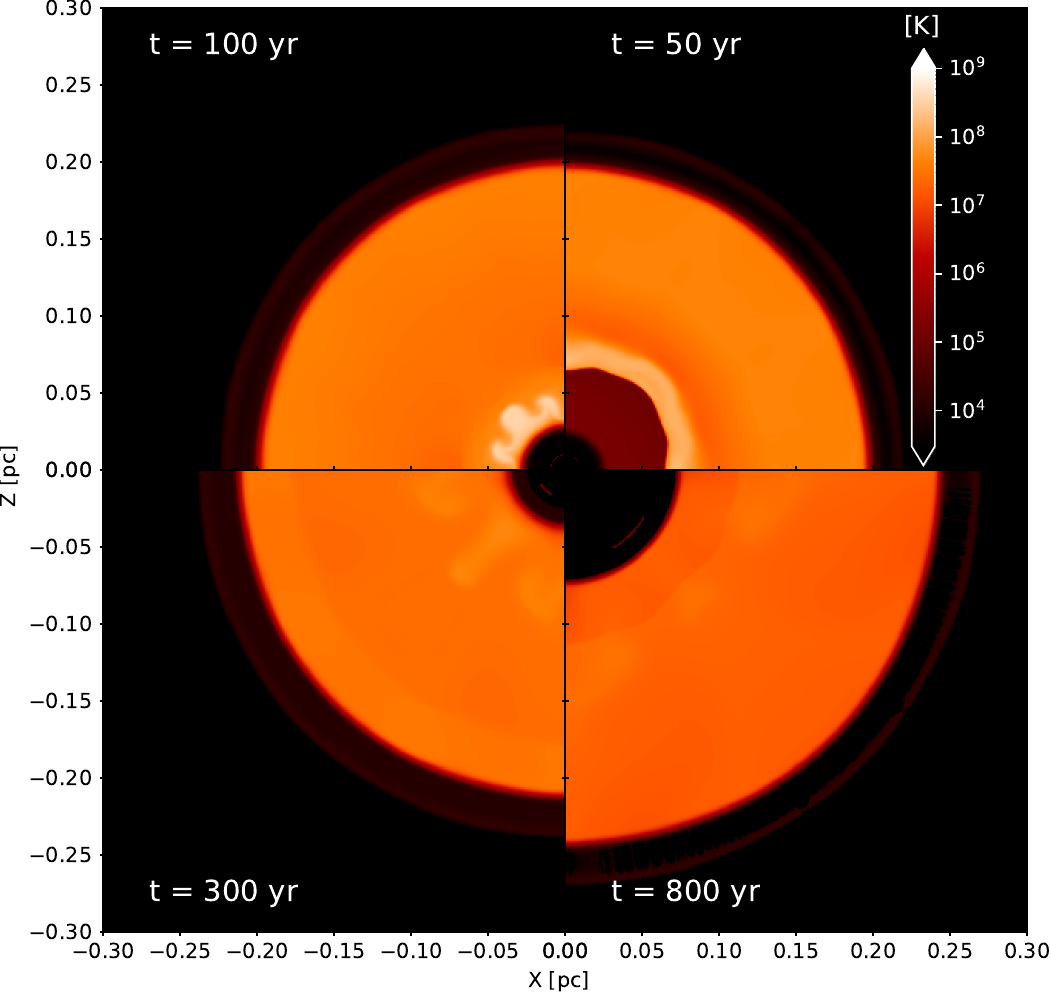}
\includegraphics[width=0.47\linewidth]{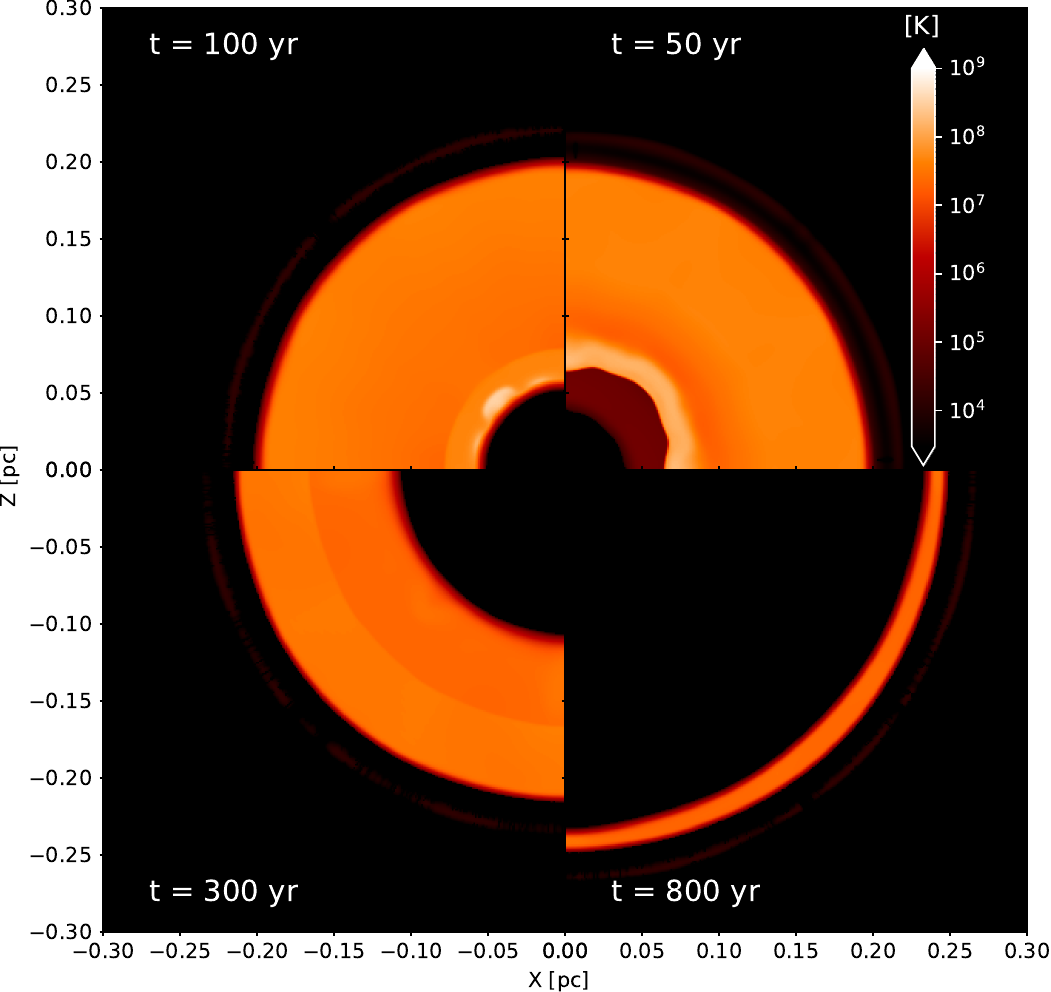}
\caption{Same as Figure~\ref{fig:rho} but for the gas temperature $T$.}
\label{fig:T}
\end{figure*}

At this point, the fast wind has carved the AGB material into a dense
shell. This interaction creates the classic adiabatically-shocked hot
bubble that fills the PN \citep[e.g.,][and references
  therein]{Toala2016}. At the same time, the strong photon flux
ionizes the material. The number density ($n$), temperature ($T$), gas
velocity ($v$) and ionization fraction ($\chi$) at this point are
illustrated in Figure~\ref{fig:initial_c}.

\subsection{The born-again phase and subsequent evolution}

\begin{figure*}
\centering
\includegraphics[width=0.47\linewidth]{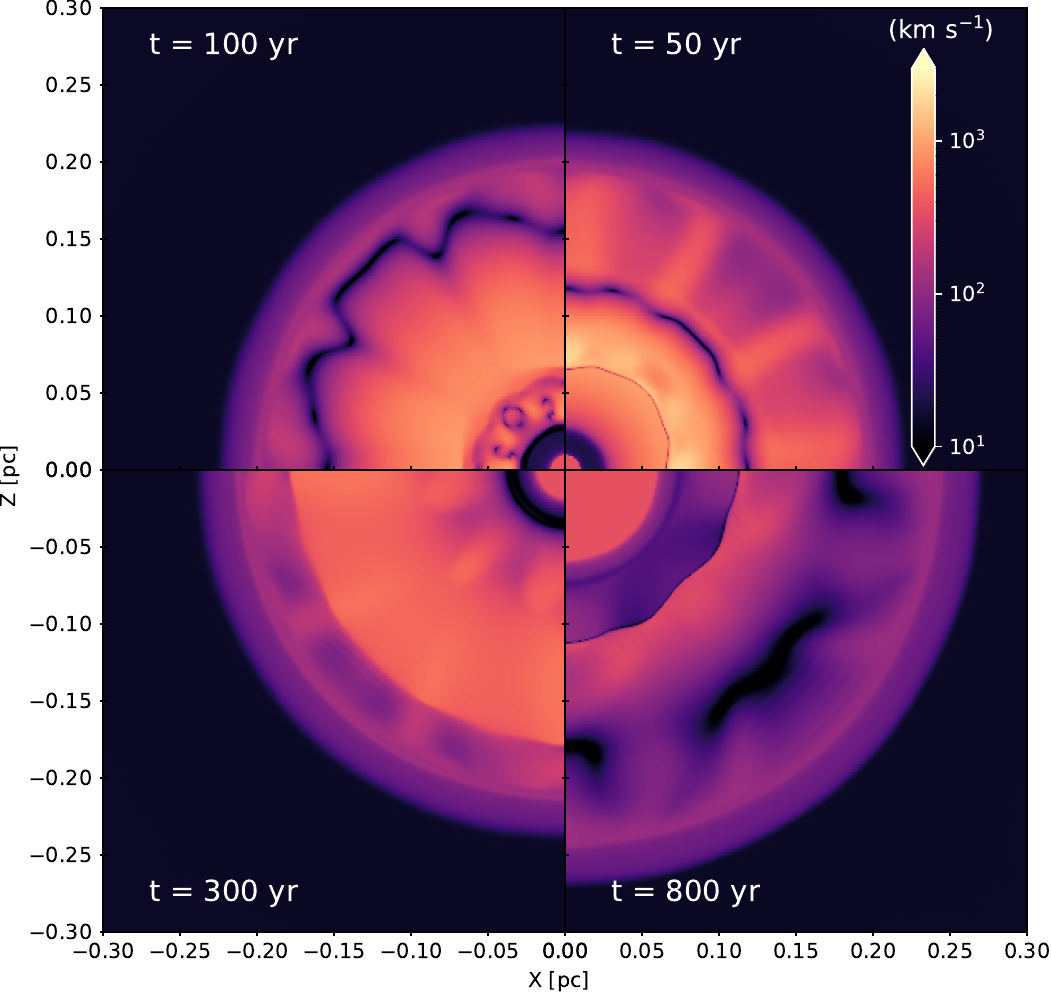}
\includegraphics[width=0.47\linewidth]{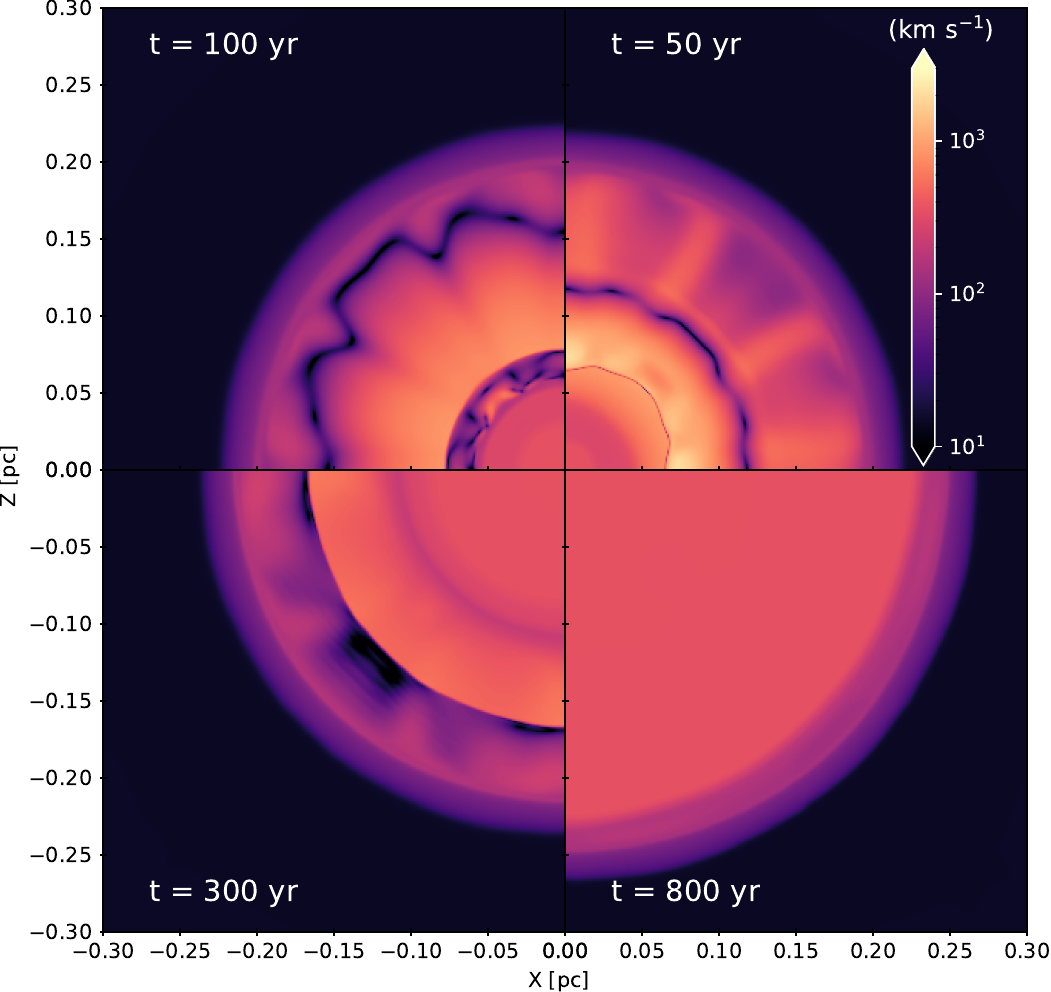}
\caption{Same as Figure~\ref{fig:rho} but for the gas velocity $v$.}
\label{fig:vel}
\end{figure*}

\begin{figure*}
\centering
\includegraphics[width=0.47\linewidth]{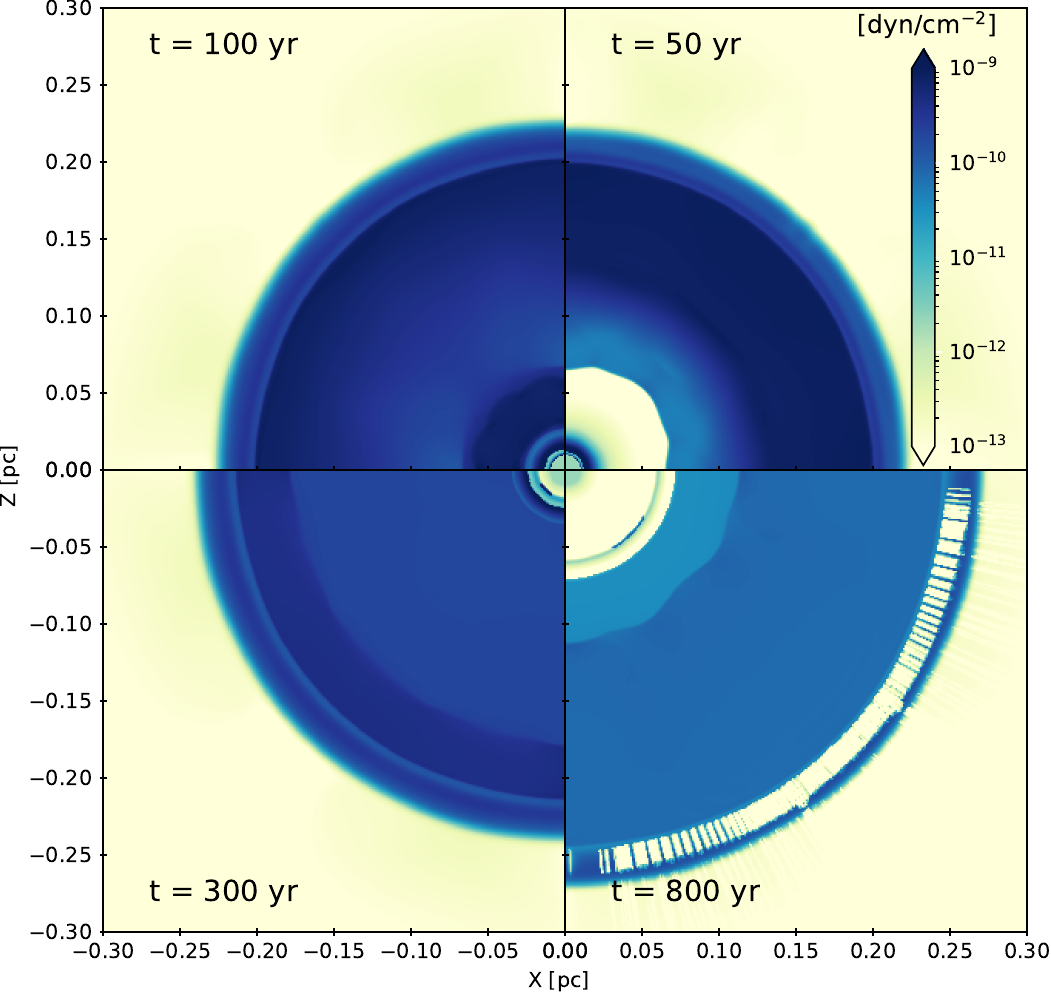}
\includegraphics[width=0.47\linewidth]{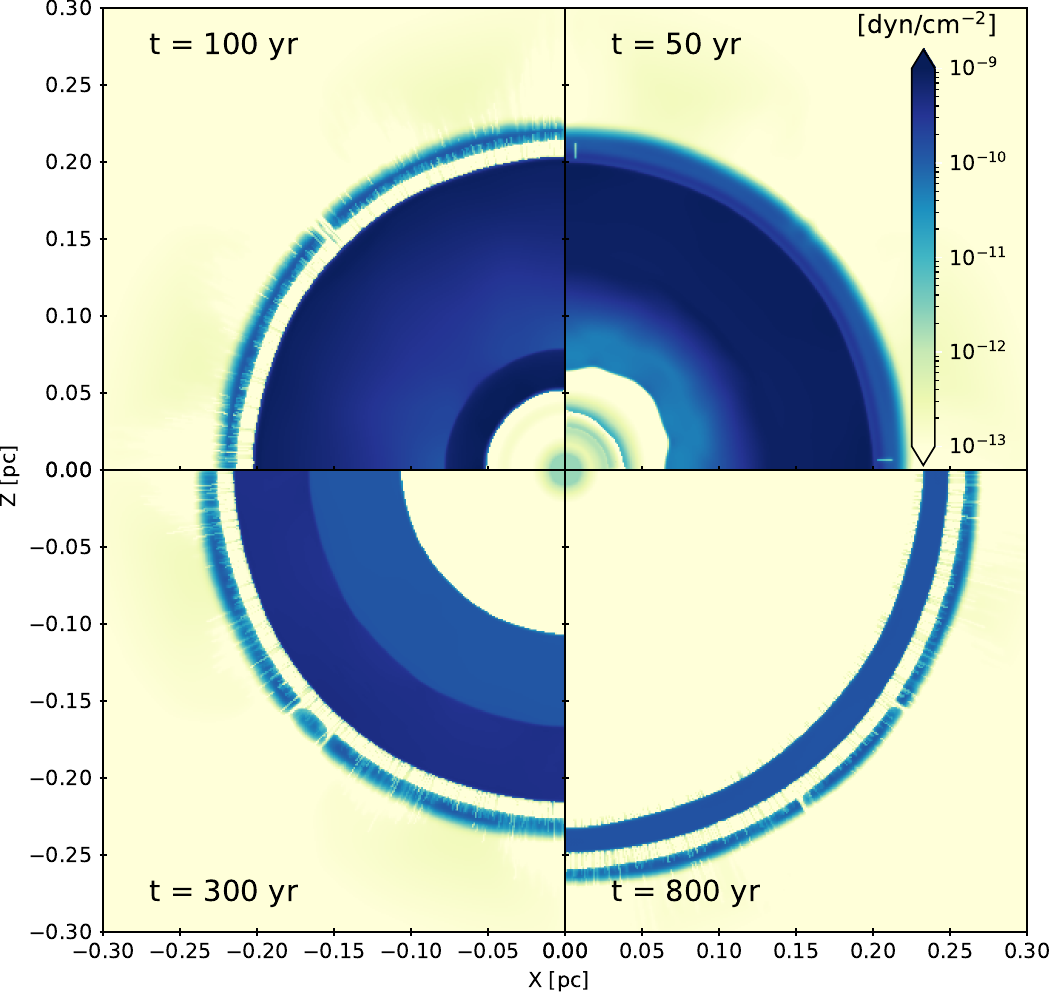}
\caption{
Same as Figure~\ref{fig:rho} but for the gas pressure $P$.
}
\label{fig:pressure}
\end{figure*}

The stellar evolution model presented for HuBi\,1 in
\citet{Guerrero2018} predicts that its CSPN experienced a mass-loss
rate during the VLTP of
$\dot{M}_\mathrm{VLTP}=7.6\times10^{-5}$~M$_\odot$~yr$^{-1}$. Following
the multi-epoch study of the Sakurai's Object, we will adopt a
duration for the VLTP of 20~yr \citep{Evans2020}. However, the
velocity of the H-poor ejected material is an unknown parameter. One
might argue that as the star goes back to the region of the AGB stars
in the Hertzsprung–Russell diagram, a similar velocity as that
reported for those kind of stars should be adopted
\citep[$\approx$20~km~s$^{-1}$; see][]{Ramstedt2020}, but the VLTP is
an explosive event in nature.

To assess both scenarios we ran two simulations. Run~A will be
performed with $v_\mathrm{VLTP1} = 20$ km~s$^{-1}$ and Run~B with
$v_\mathrm{VLTP2} = 300$ km~s$^{-1}$, similar to what is observed for
the H-poor ejecta in HuBi\,1 \citep{Rechy2020}. No ionizing photon
flux will be considered during this phase. Figure~\ref{fig:hist}
illustrates the variations of the mass-loss rate and velocity during
this phase (between $t_2$ and $t_3$) in comparison with the previous
phases. As a result of the high-mass loss rate during the VLTP phase
and its short duration we will create a dense shell surrounding the
CSPN with an extension of $\lesssim$0.02~pc in radius.

A final post-VLTP (pVLTP) phase will be modelled by adopting the
stellar wind parameters currently exhibited by the CSPN of HuBi\,1
reported in \citet{Guerrero2018}.  A pVLTP wind velocity of
$v_\mathrm{pVLTP} = 360$ km~s$^{-1}$ with a mass-loss rate of
$\dot{M}_\mathrm{pVLTP}=8\times10^{-7}$~M$_{\odot}$~yr$^{-1}$ for both
Run~A and B.  This wind is expected to sweep the VLTP ejecta creating
a dense inner shell and giving rise to the double shell morphology.
The ionizing photon flux of $10^{44}$~s$^{-1}$ reported by
\citet{Guerrero2018} will be adopted for this last phase.

Figure~\ref{fig:hist} illustrate the variations of the mass-loss rate
and stellar wind velocity from the AGB phase to the pVLTP for the two
simulations which corresponds to $t>t_3$.

\begin{figure*}
\centering
\includegraphics[width=0.45\linewidth]{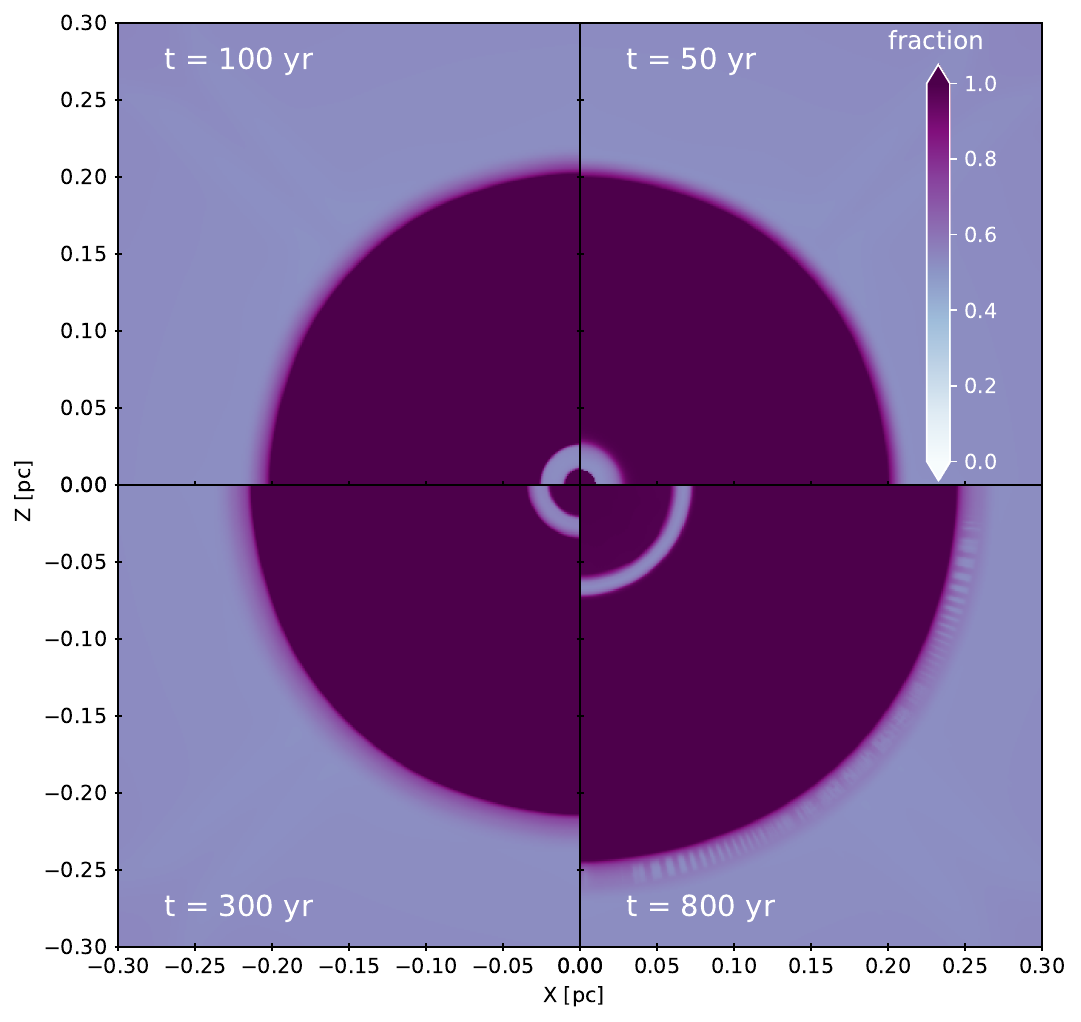}
\includegraphics[width=0.45\linewidth]{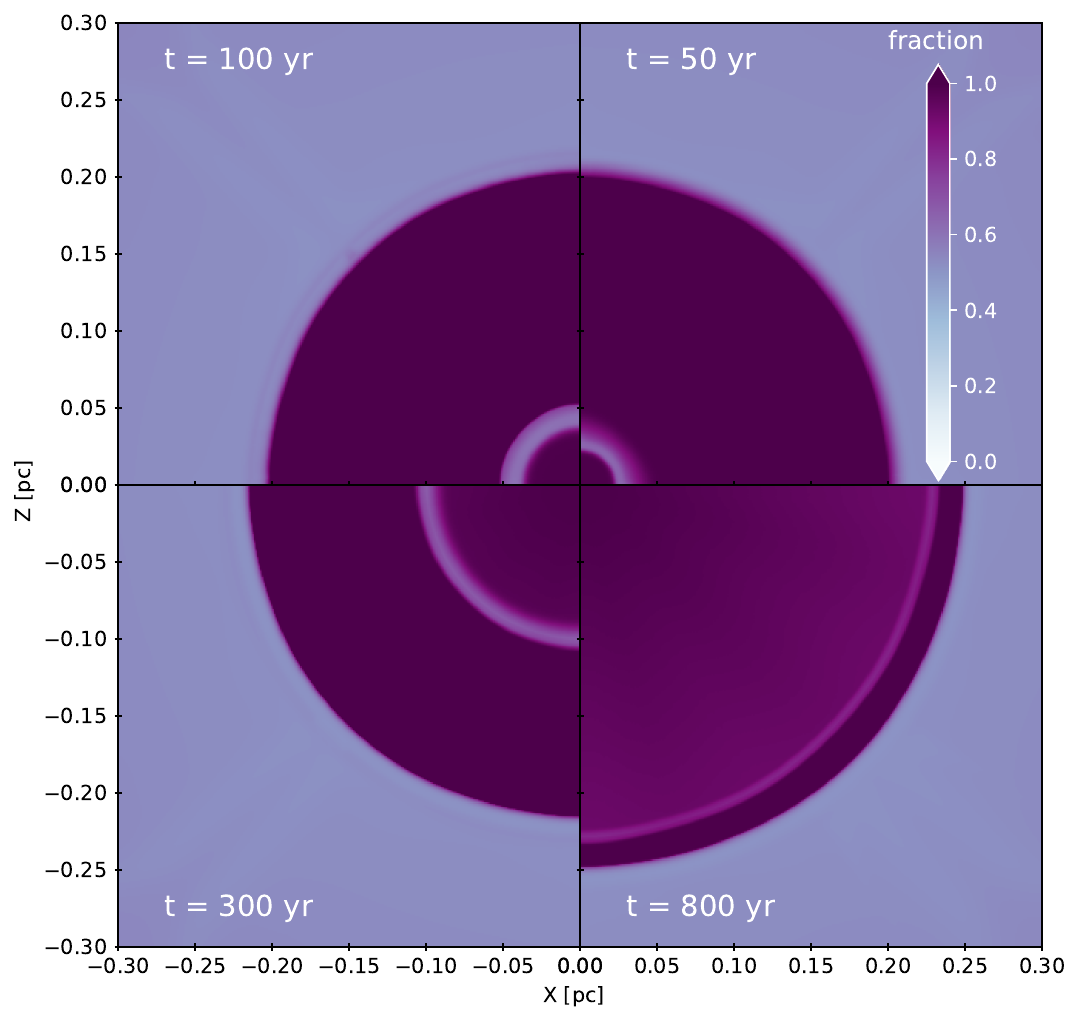}
\caption{
Same as Figure~\ref{fig:rho} but for the ionization fraction $\chi$.
}
\label{fig:ionization}
\end{figure*}

\section{Results}

Figures~\ref{fig:rho}, \ref{fig:T}, \ref{fig:vel}, \ref{fig:pressure}
and \ref{fig:ionization} show the evolution with time of $n$, $T$,
$v$, $P$ and $\chi$ of the gas for the two simulations presented here,
Run~A and B.  For simplicity, $t$=0 has been set to the end of the
VLTP $t=t_3$ in all subpanels of these figures.  These subpanels thus
represent snapshots at different times after the onset of the pVLTP.

Figure~\ref{fig:rho} and \ref{fig:T} show that the dense VLTP material
expands into the low-density, hot bubble created by the previously
fast $v_\mathrm{PN}$ wind.  By 50~yr of evolution into the pVLTP
phase, the two simulations show the formation of a dense shell with
radius between 0.02 and 0.03~pc (Fig.~\ref{fig:rho}).

It is important to note that due to the sudden variation of the
stellar wind parameters and the ionizing photon flux, the inner edge
of the hot bubble experiences noticeable changes.  In particular, the
ram pressure of the wind in the VLTP is not as high as that of the
fast wind that fed the previous PN phase, which created the hot bubble
(see Fig.~\ref{fig:initial_c} top left panel).  As a result, the hot
bubble falls back to smaller radii when the star evolves into the VLTP
phase, creating instabilities that develop with time. Such effects are
more evident in the gas velocity and pressure (Fig.~\ref{fig:vel} and
\ref{fig:pressure}), with turbulent structures appearing at 100~yr
after the pVLTP evolution, and are still noticeable after 300~yr and
to some extent in the most evolved panel at 800~yr of Run~A.  There is
no apparent effect in the ionization fraction of these turbulent
structures (see Fig.~\ref{fig:ionization}), implying that they are
completely ionized in our simulations.

The relatively small variation in velocity between the VLTP and pVLTP
phases in both simulations is not enough to produce the hydrodynamic
instabilities (e.g., Rayleigh-Taylor) reported in other works
\citep{Stute2006,Toala2016}.  As a consequence, our numerical results
show the expansion of a smoothed shell expanding inside the old PN.

\begin{figure*}
\centering
\includegraphics[width=0.75\linewidth]{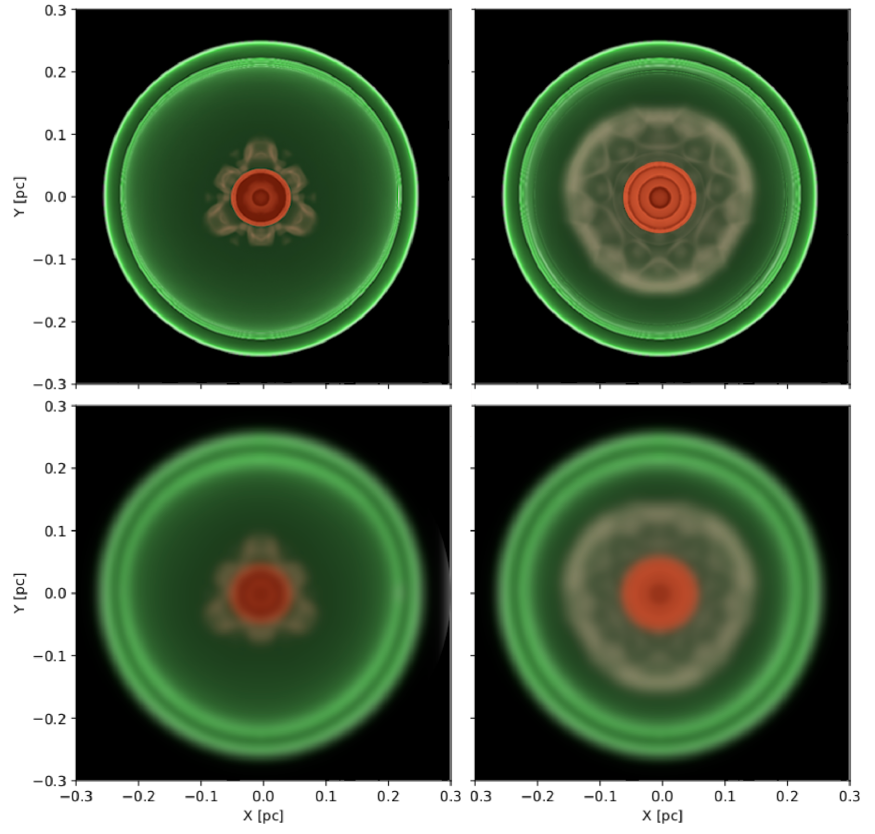}
\caption{Synthetic nebular images obtained by integrating through the
  $x-y$ plane ($z=0$) for the two simulations presented in this
  work. The images were produced at the time at which the inner shell
  reaches 0.05~pc, i.e., 600~yr after the onset of the pVLTP phase for
  Run~A (left panel) and 100~yr for Run~B (right panel). The bottom
  panels show the same images at the spatial resolution of the NOT
  images ($\sim1''$).}
\label{fig:rendered}
\end{figure*}

Finally volume density renderings were created to mimic nebular images
for both simulations at an integration time when the inner shell has a
radius of 0.05~pc.  The top-left panel of Figure~\ref{fig:rendered}
shows the case of Run~A at 600 yr after the pVLTP evolution, while the
top right panel shows Run~B at 100~yr of evolution.  The rendered
images were produced with the {\sc yt} software \citep{Turk2011} that
allows to create images by casting rays through the 3D volume and
integrating the radiation transfer equations with a transfer function
that can be selected or modified by the user.  Typically, the transfer
function is chosen with transparency and/or colors that depend on the
value of the field that is being rendered.  The images presented in
Fig.~\ref{fig:rendered} use an opacity that varies linearly with
density (denser is more opaque), considering only the densest regions,
which correspond to number densities between $0.1$ and
$1000~\mathrm{cm^{-3}}$.

We use a brown (low density) to green (high density) tinge for the
material injected before the born-again phase, and superimposed it
with a red tinge and the same opacity law for the material injected
after the born-again phase.  We must note that is not possible to make
a quantitative comparison with the observations, as the emission
detected arises from recombination and forbidden lines, which are
proportional to the density squared and depend on the ionization stage
of each particular element.  This would require a detailed description
of the ionization structure that is beyond the scope of the present
paper.  To allow a fairer comparison with the available NOT images, a
Gaussian filter was apply to reduce the spatial resolution of the
rendered images to $\sim1''$. The synthetic images in the bottom
panels of Figure~\ref{fig:rendered} reproduce the double shell
morphology of HuBi\,1, revealing an additional clumpy intermediate
structure which is present in Figure~\ref{fig:opt}, particularly in
the contrast enhanced [N~{\sc ii}] image in its right panel.

\section{Discussion}

The born-again phase is one of the most unknown phases of stellar
evolution.  Its duration seems to depend on different factors such as
mixing and diffusion as discussed in \citet{MB2006}.  However, some
efforts assessing the mass lost during this phase and the velocity of
the ejected H-poor material have been presented in the literature
\citep[see][and references there]{Guerrero2018,Toala2021}.  This can
certainly help driving theoretical results.

The results of our simulations for HuBi\,1 suggest that it is more
accurate to assume that the H-poor material was ejected inside the old
PN in an explosive event with velocities close as those currently
observed.  Run~B predicts that if this is the case, the material
ejected during the VLTP in HuBi\,1 must have occurred $\sim$100~yr,
which is rather consistent with the kinematical age estimated by
\citet{Rechy2020}.  Run~B also suggests that the small difference in
velocity between the VLTP and the pVLTP does not allow the shell to
experience the formation of hydrodynamical instabilities, in
particular, Rayleigh-Taylor.  As a result, the inner shell of HuBi\,1
appears to be a smooth shell-like structure.  Our simulations predict
that the inner shell will not develop hydrodynamic instabilities
capable of disrupting it.  In contrast, the simulations presented in
\citet{Fang2014} for A\,30 and A\,78 suggest that the combination of
these physical processes facilitate the disruption of their dense VLTP
shells.  This simulation took into account the fast evolution of their
central early-type [WC] stars, which currently exhibit stellar wind
velocity of $\approx$3000~km~s$^{-1}$ and high ionizing photon fluxes.

We have shown that the relatively low ionization photon flux of
10$^{44}$~s$^{-1}$ suggested from the stellar atmosphere modelling of
the CSPN of HuBi\,1 is not able to completely ionize the born-again
inner shell.  In our simulations the H-poor ejecta has a ionization
fraction close to 0.5.  This strengthens \citet{Guerrero2018}'s
suggestion that the emission from this shell must come from shock
physics.  A careful analysis of the optical emission spectrum will
address this issue (Montoro-Molina et al., in preparation).

\begin{figure}
\centering
\includegraphics[width=\linewidth]{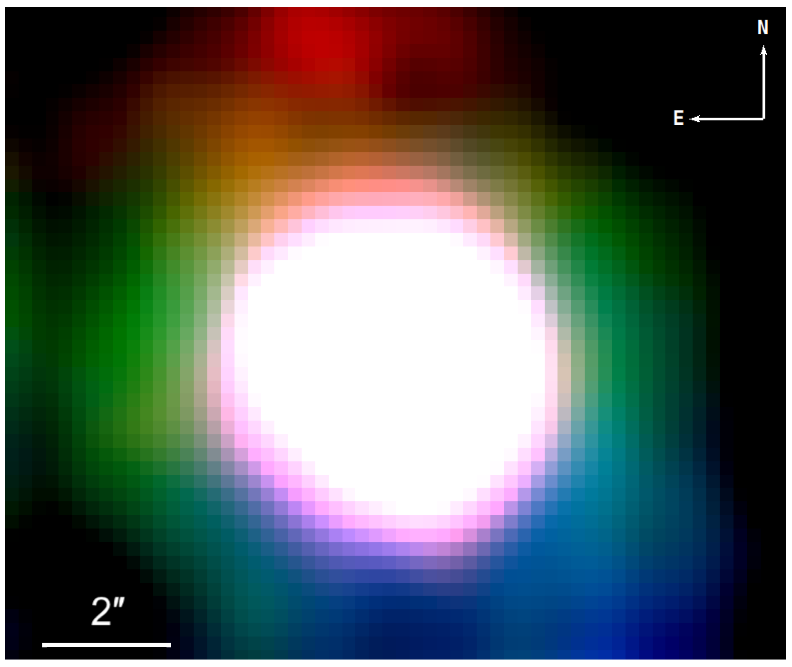}
\caption{[N\,{\sc ii}] emission from the GTC MEGARA observations of
  HuBi\,1. The green colour represents the [N\,{\sc ii}] emission
  centered on the systemic velocity of HuBi\,1. The blue and red
  colour show the integrated velocity in the [$-$55:$-$45]~km~s$^{-1}$
  and [57:66]~km~s$^{-1}$, respectively. The inner shell of HuBi\,1
  appears saturated in white.}
\label{fig:NII_prof}
\end{figure}

Figure~\ref{fig:rendered} suggests that the structures detected
between 2~arcsec$<r<$5~arcsec in the [N\,{\sc ii}] image surrounding
the born-again ejecta of HuBi\,1 (see Fig.~\ref{fig:opt} right panel)
have been formed as a result of the sudden and large variation of the
stellar wind parameters of its CSPN.  These turbulent structure
appears to be completely ionized, unlike the inner shell.  Indeed
Figure~1 in \citet[][]{Guerrero2018} suggests that this is the case.
To explore the velocity structure of this emission, we have examined
the GTC MEGARA integral field spectroscopic data published recently by
our team in \citet{Rechy2020}.  The results are illustrated in
Figure~\ref{fig:NII_prof}, a colour-composite picture of HuBi\,1 in
the [N\,{\sc ii}] $\lambda\lambda$6548,6584 emission lines where the
green colour corresponds to [N\,{\sc ii}] emission at the systemic
velocity of HuBi\,1, the blue colour to the approaching structure in
the systemic velocity range from $-55$ to $-45$ km~s$^{-1}$, and the
red colour to the receding structure in the systemic velocity range
from $+57$ to $+66$ km~s$^{-1}$. Other emission lines detected in the
MEGARA data cube, such H$\alpha$ and [S\,{\sc ii}], show similar
velocity structures, but the former suffers from thermal broadening,
while the later has a smaller signal-to-noise level than the [N\,{\sc
    ii}] image presented here.

Figure~\ref{fig:NII_prof} shows that the structure surrounding the
inner shell of HuBi\,1 has a somewhat complex velocity structure.
Some emission at the systemic velocity of HuBi\,1 might have formed as
illustrated by our simulations, gas falling back due to the reduction
in ram pressure generating instabilities in the ionized
structure. However, there seems to be a bipolar structure not reported
before in HuBi\,1. This bipolar structure does not appear to be
collimated, but extended at a certain point. A detailed analysis of
the velocity of this structure using our available GTC MEGARA
observations is under preparation (Montoro-Molina et al., in
preparation).

\subsection{Consequences for other born-again PNe}

The simulation of Run~A reaches a radius for the inner shell of
0.05~pc after 600~yr of evolution, which is notably different to the
age of $\simeq$200 yr proposed by \citet{Rechy2020}.  The model can
neither reproduce the reported expansion velocities of the inner shell
in HuBi\,1, regardless of the injection of a pVLTP wind 20 times
faster, which can not provide sufficient kinetic energy to accelerate
the shell up to the observed velocities \citep[$\sim$300
  km~s$^{-1}$;][]{Rechy2020}, whereas the momentum provided by the
radiation pressure is negligible.  Still, it is appropriate to discuss
the numerical results of Run~A as a slow expanding VLTP wind might
have been the case for A\,30 and A78, where after $\sim$1000~yr of
evolution in the born-again phase dense knots and filaments are still
located close to their CSPN with expansion velocities
$\lesssim$50~km~s$^{-1}$ \citep{Meaburn1996,Meaburn1998}.

The 2D Radiation-hydrodynamic simulations presented in
\citet{Fang2014} showed that it is possible to reproduce the
morphology of the H-deficient clumps and filaments in A\,30 and A\,78
only if the velocity during the VLTP phase in these objects was
$\sim$20~km~s$^{-1}$, followed by a fast subsequent evolution of the
stellar wind parameters reaching terminal velocities as high as
$\gtrsim$3000~km~s$^{-1}$, which is what is currently reported from
these sources \citep[see][]{Guerrero2012,Toala2015}.  The interactions
between the fast pVLTP and slow VLTP winds are dominated by
Rayleigh-Taylor instabilities that create a pattern of slow clumps and
filaments lagging close to the CSPN, whilst evaporated material can
reach velocities of a few times 100 km~s$^{-1}$ with distances almost
reaching the outer H-rich PNe.

The notable differences between the most evolved born-again PNe
discovered so far and the {\it inside out} PN HuBi\,1 might suggest
that the explosive VLTP might have had different injection
energies. The kinetic energy imprinted in the H-deficient material
ejected inside the old PN should be directly related to the He mass
ignited during the VLTP (born-again) event. Assuming that the total
mass ejected in these born-again PNe is the same, the kinetic energy
of the H-deficient material in HuBi\,1 is $\gtrsim$30 times larger
than the slowly moving ($\sim$50~km~s$^{-1}$) dense clumps in A\,30
and A\,78. The later suggest that the thermonuclear conditions of the
VLTP were quite different between these systems.

We are currently preparing a grid of stellar evolution models
accounting for different parameters such as initial mass, rotation,
metallicity, mixing length (Rodr\'{i}guez-Gonz\'{a}lez et al.\ 2021,
in prep.) using the Modules for Experiments of Stellar Astrophysics
\citep[{\sc mesa};][]{Paxton2011}.  These will help us to assess
possible different conditions occurring during the VLTP.  Furthermore,
increasing the number of identified born-again PNe is most needed to
shed some light into this short but unique and physically complex
evolution phase of low-mass stars.

\section{Summary} 
\label{sec:summary}

We presented the first 3D radiation-hydrodynamic numerical simulations
of the formation and evolution of a born-again PNe, with particular
application to the inside-out PN HuBi\,1.  We adapted the stellar wind
parameters and ionization photon flux reported by \citet{Guerrero2018}
for the CSPN of HuBi\,1 to produce tailored numerical simulations.
Since the velocity of the H-poor material ejected during the VLTP
phase is unknown, two different simulations were presented to explore
its effects: in Run~A we adopted an expansion velocity
$v_\mathrm{VLTP1}$=20 km~s$^{-1}$, similar to that is reported for AGB
stars, while in Run~B we adopted a higher velocity
$v_\mathrm{VLTP2}$=300 km~s$^{-1}$, consistent with that reported from
optical observations in HuBi\,1.  Our findings can be summarized as
follows:
\begin{itemize}

\item 
Our explosive case, Run~B, makes a good job reproducing the
morphological features in HuBi\,1.  These simulations predict that the
inner shell of HuBi\,1 was formed as a result of the born-again event
which occurred about 100~yr ago, consistent with kinematic estimations
from GTC MEGARA observations.  Slower ejections can not imprint the
necessary kinematic energy to accelerate the H-deficient material to
the observed velocity of 300~km~s$^{-1}$.

\item 
Our simulations show that the small variation in velocity between the
VLPT and the pVLTP material so far observed will not produce
instabilities that break the inner shell in contrast to the more
evolved born-again PNe A\,30 and A\,78. This produces a smooth inner
shell consistent with that seeing in optical observations.

\item 
The extreme changes experienced by the CSPN of HuBi\,1 are obviously
responsible of the double shell morphology seen in optical
observations (Fig.~\ref{fig:opt} and \ref{fig:NII_prof}).  Moreover,
the variation in the stellar wind parameters diminishs dramatically
the wind's ram pressure, producing noticeable changes to the
adiabatically shocked hot region originally created at the first PN
phase.  The hot bubble falls back by the time the star enters the VLTP
phase, producing turbulent structures which are observable as clumps
and filaments of ionized material at intermediate regions between the
two shells.  We propose this is the origin of the structures detected
in the [N\,{\sc ii}] image of HuBi\,1 in the intermediate regions
between 2 and 5~arcsec.
  
\item 
Our simulations demonstrate that the current photon flux of
10$^{44}$~s$^{-1}$ is not capable of producing the complete
photoionization of the inner shell of HuBi\,1.  This result
strengthens the suggestion of \citet{Guerrero2018} that this structure
is dominated by shocks.

\item 
We suggest that the explosive VLTP in HuBi\,1 might have been at least
30 times more energetic than that in the born-again PNe A\,30 and
A\,78. Dense clumps in A\,30 and A\,78 have been detected very close
to the CSPN with velocities as low as $\lesssim$50~km~s$^{-1}$ which
have survived for about 1000~yr.  Such differences puts under scrutiny
the physics involved in producing the VLTP and suggests a wealth of
initial conditions in this scenario.

\end{itemize}


\section*{Acknowledgements}

JAT thanks funding by Fundación Marcos Moshinsky (Mexico) and the
Direcci\'{o}nn General de Asuntos del Personal Acad\'{e}mico (DGAPA)
of the Universidad Nacional Aut\'{o}noma de M\'{e}xico (UNAM) project
IA100720. V.L. gratefully acknowledges support from the \mbox{CONACyT}
Research Fellowship program.  BMM and MAG acknowledge support of the
Spanish Ministerio de Ciencia, Innovaci\'on y Universidades (MCIU)
grant PGC2018-102184-B-I00.  AE acknowledges support from DGAPA-PAPIIT
(UNAM) grant IN 109518.  This work has made extensive use of NASA's
Astrophysics Data System.

\section*{Data availability}
The data underlying this work are available in the article. Our
results will be shared on reasonable request to the first author.


\end{document}